\documentclass[trackchanges,resetfootnote]{aastex7}

\newcommand{\astroquery}{\texttt{astroquery}}

\begin{document}

\title{An in-depth study of brown dwarfs with TESS}

\correspondingauthor{Frédéric Marcadon}
\email{fmarcadon@camk.edu.pl}

\author[0000-0002-7606-7733]{Frédéric Marcadon}
\affiliation{Nicolaus Copernicus Astronomical Center, Polish Academy of Sciences, ul.\ Bartycka 18, 00-716 Warszawa, Poland}
\affiliation{Department of Astrophysics and Planetary Sciences, Villanova University, 800 East Lancaster Avenue, Villanova, PA 19085, USA}
\email{fmarcadon@camk.edu.pl}  

\author[0000-0002-1913-0281]{Andrej Pr\v{s}a}
\affiliation{Department of Astrophysics and Planetary Sciences, Villanova University, 800 East Lancaster Avenue, Villanova, PA 19085, USA}
\email{andrej.prsa@villanova.edu}


\begin{abstract}

The existence of a deficit of brown dwarfs (BDs) in close orbit around main-sequence stars is one of the most intriguing questions in stellar physics. This so-called BD desert may result from the transition between two different dominant formation processes occurring for different mass regimes. While the BD mass derived from radial-velocity measurements helps confirm the nature of the analyzed objects, the BD radius obtained from transits is important to better constrain the BD age, as BDs are believed to contract with age. Such objects with well-constrained parameters, although in small number, are of prime interest for deeper investigations of BD structure and chemical composition. The present document aims at presenting the first results of a search for BD transits among a sample of approximately 3300 host star candidates observed by the Transiting Exoplanet Survey Satellite during Cycle~6.

\end{abstract}

\keywords{\uat{Brown dwarfs}{185} --- \uat{Transit photometry}{1709}}




\section{Introduction} 

Brown dwarfs (BDs) are among the least studied objects in the field of star formation. The detection of these objects has long remained a challenge due to their low intrinsic luminosity. Though their existence was hypothesized in the 1960s, the first BD, Teide~1, was discovered only in 1995 in the Pleiades star cluster \citep{1995Nature.377..129R}. BDs are commonly defined as substellar objects with masses between 13 and 80$\,M_{\rm Jup}$, that is between the deuterium and hydrogen burning boundaries separating BDs from giant planets and very low-mass stars. However, these boundaries are not well defined as they depend on the chemical composition of the BD \citep{2014AJ....147...94D} and possibly on its formation history. Low-mass BDs may form like giant planets by core accretion \citep{1996Icar..124...62P} or disk instability \citep{1997Sci...276.1836B} whereas high-mass BDs may form like stars by collapse and fragmentation of molecular clouds \citep{2004ApJ...617..559P}, either as a single object or in a binary system. In the latter case, recent studies have confirmed the presence of a BD desert (e.g., \citealt{2023A&A...680A..16U}), corresponding to a deficit of BDs in close orbit ($\lesssim\:\!$5$\,$au) around main-sequence stars. Although the origin of the BD desert remains currently unclear, part of the answer may lie in how these objects form and evolve. 

In order to understand BD formation and evolution in more detail, it is necessary to explore the diversity of BD population properties. In particular, over the past 7 years, the Transiting Exoplanet Survey Satellite (TESS) allowed the detection of a steadily growing number of transiting BDs with well-determined radii and masses \citep{2025A&A...696A..44B}. As a nearly all-sky transit survey, TESS can monitor a large number of transiting BD candidates. A precise determination of BD radii from transits is then crucial to test how well evolutionary models predict the radius evolution of BDs. It is thus expected that the detection and characterization of new transiting BDs will help us to better constrain the BD population properties in the era of TESS.



\section{Target selection}

We selected a sample of 5098 targets observed by TESS during Cycle~6 as part of our Guest Investigator program
G06058 (PI: F.~Marcadon) and identified as candidate or confirmed substellar objects in the SIMBAD database through a cross-identification of objects from catalogs and articles. This sample includes 15 of the $\sim$50 confirmed transiting BDs reported in the literature \citep{2025A&A...696A..44B}. We ordered these targets according to their potential for asteroseismology and transiting BD searches. For each target, we derived the value of the frequency of maximum oscillation power from the scaling relation $\nu_{\rm max} \propto g / \sqrt{T_{\rm eff}}$, where the values of $g$ and $T_{\rm eff}$ were taken from the TIC catalog \citep{2019AJ....158..138S}. Based on the values of $\nu_{\rm max}$ and $T_{\rm mag}$, as well as on the number of observed sectors, we divided our sample into three target prioritization levels.  The lowest priority (1862 objects) was given to targets with a number of observed TESS sectors $N_{\rm sectors} \leq 3$, while the highest and medium priorities (44 and 3192 objects, respectively) were assigned to targets with $N_{\rm sectors} \geq  4$. In particular, the highest priority targets correspond to the brightest stars ($T_{\rm mag} < 10$) in our sample for which we expect to detect solar-like oscillations with $\nu_{\rm max}$ below the TESS 2 minute Nyquist frequency of 4167$\,\mu$Hz.



\section{Analysis and results}

We first downloaded the 2 minute cadence data of $\sim$3300 targets from Sectors 14 to 78 using the \astroquery{} package \citep{2019AJ....157...98G}. In our analysis, we chose to use the Pre-search Data Conditioning Simple Aperture Photometry (PDCSAP) light curves \citep{2012PASP..124.1000S,2012PASP..124..985S,2014PASP..126..100S}, which have been corrected for instrumental effects and contamination by nearby stars. Then, we normalized the light curves by dividing by the median flux value of each sector. No outlier rejection was applied to the light curves, to avoid removing possible transit signals. In a second step, we performed a transit search using the Transit Least Squares (TLS) algorithm \citep{2019A&A...623A..39H}, which proved to be more efficient than the Box Least Squares (BLS) method \citep{2002A&A...391..369K}, with the default settings. For each target, TLS creates a linear frequency grid covering a range between $\sim$0.6 days and half the duration of the observations. We flagged the objects with a false alarm probability smaller than 0.01\% and phase-folded the light curves according to the transit ephemerides obtained with TLS. 

After visual inspection of the phase-folded light curves, we identified 366 objects with significant periodic signals ranging from hours to tens of days. The ephemerides of these objects are given in Table~\ref{tab:tic}, along with the signal detection efficiency (SDE) values. Among them, we found a variety of variable sources such as transiting BD candidates, transiting hot Jupiters, and eclipsing binaries (EBs). We are currently in the process of uploading the light curves of the 366 variable candidates to the Villanova TESS EB catalog\footnote{See \url{http://tessebs.villanova.edu/}.}. The portal has been operational since 2022, maintained by our research group, and the service is available to the public from the moment of data ingestion.



\section{Future prospects}

In this work, we conducted a systematic search for new transiting BD systems using TESS 2 minute cadence photometry, which resulted in the detection of 366 variable candidates. Future prospects for analyzing these objects are discussed below:
\begin{enumerate}
    \item{Over the past decade, asteroseismology has matured into a powerful tool to precisely determine the fundamental properties of stars, leading to the characterization of $\sim$110 exoplanet host stars \citep{2022AJ....163...79H}. Thus, in future work, we plan to apply the method developed by \cite{2009A&A...508..877M} to the residual TESS light curves in order to search for potential solar-like oscillations} in our sample of transiting BD host star candidates.
    \item As a byproduct of our search for BD transits, we also identified a number of EBs. These targets have benefited from TESS 2 minute cadence observations, which are ideal for eclipse timing variation (ETV) studies (see, e.g., \citealt{2024ApJ...976..242M}). In particular, the ETV analysis of these EBs should allow us to detect new compact hierarchical triples \citep{2025A&A...695A.209B} and substellar circumbinary companions \citep{2025arXiv250917011M}.
    \item In a continuation of this work, we plan to perform the transit timing variation (TTV) analysis of the detected transiting hot Jupiters. Three competing scenarios attempt to explain the formation of hot Jupiters: in situ formation, disk migration, and high-eccentricity migration. However, the near absence of observed nearby planetary companions in hot Jupiter systems seems to support the high-eccentricity migration scenario, as suggested by \citet{2024A&A...692A.254H} and \citet{2024ApJS..275...32Z}. We aim to address this question by searching for additional planetary companions in our hot Jupiter sample using the TTV method. 
\end{enumerate}



\begin{acknowledgments}
This paper includes data collected by the TESS mission, which are publicly available from MAST. Funding for the TESS mission is provided by NASA’s Science Mission directorate.

We gratefully acknowledge support from the NASA TESS Guest Investigator grant 80NSSC24K0498 (PI: F.~Marcadon).
\end{acknowledgments}





%
\facilities{TESS.}

\software{\astroquery{} \citep{2019AJ....157...98G} and
    TLS \citep{2019A&A...623A..39H}.
          }



\bibliography{sample7}{}

\begin{thebibliography}{}
\expandafter\ifx\csname natexlab\endcsname\relax\def\natexlab#1{#1}\fi
\providecommand{\url}[1]{\href{#1}{#1}}
\providecommand{\dodoi}[1]{doi:~\href{http://doi.org/#1}{\nolinkurl{#1}}}
\providecommand{\doeprint}[1]{\href{http://ascl.net/#1}{\nolinkurl{http://ascl.net/#1}}}
\providecommand{\doarXiv}[1]{\href{https://arxiv.org/abs/#1}{\nolinkurl{https://arxiv.org/abs/#1}}}

\bibitem[{K. {Barkaoui} {et~al.}(2025){Barkaoui}, {Sebastian},
  {Z{\'u}{\~n}iga-Fern{\'a}ndez}, {Triaud}, {Rackham}, {Burgasser},
  {Carmichael}, {Gillon}, {Theissen}, {Softich}, {Rojas-Ayala}, {Srdoc},
  {Soubkiou}, {Fukui}, {Timmermans}, {Stalport}, {Burdanov}, {Ciardi},
  {Collins}, {Davis}, {Davoudi}, {de Wit}, {Demory}, {Deveny}, {Dransfield},
  {Ducrot}, {Florian}, {Gan}, {G{\'o}mez Maqueo Chew}, {Hooton}, {Howell},
  {Jenkins}, {Littlefield}, {Mart{\'\i}n}, {Murgas}, {Niraula}, {Palle},
  {Pedersen}, {Pozuelos}, {Queloz}, {Ricker}, {Schwarz}, {Seager}, {Shporer},
  {Scott}, {Stockdale}, \& {Winn}}]{2025A&A...696A..44B}
{Barkaoui}, K., {Sebastian}, D., {Z{\'u}{\~n}iga-Fern{\'a}ndez}, S., {et~al.}
  2025, \bibinfo{title}{{TOI-6508 b: A massive transiting brown dwarf orbiting
  a low-mass star},} \aap, 696, A44, \dodoi{10.1051/0004-6361/202453508}

\bibitem[{T. {Borkovits} {et~al.}(2025){Borkovits}, {Rappaport}, {Mitnyan},
  {B{\'\i}r{\'o}}, {Cs{\'a}nyi}, {Forg{\'a}cs-Dajka}, {Forr{\'o}}, {Hajdu},
  {Seli}, {Sztakovics}, {G{\"o}bly{\"o}s}, \& {P{\'a}l}}]{2025A&A...695A.209B}
{Borkovits}, T., {Rappaport}, S.~A., {Mitnyan}, T., {et~al.} 2025,
  \bibinfo{title}{{Then and now: A new look at the eclipse timing variations of
  hierarchical triple star candidates in the primordial Kepler field, revisited
  by TESS},} \aap, 695, A209, \dodoi{10.1051/0004-6361/202453616}

\bibitem[{A.~P. {Boss}(1997){Boss}}]{1997Sci...276.1836B}
{Boss}, A.~P. 1997, \bibinfo{title}{{Giant planet formation by gravitational
  instability.},} Science, 276, 1836, \dodoi{10.1126/science.276.5320.1836}

\bibitem[{S.~B. {Dieterich} {et~al.}(2014){Dieterich}, {Henry}, {Jao},
  {Winters}, {Hosey}, {Riedel}, \& {Subasavage}}]{2014AJ....147...94D}
{Dieterich}, S.~B., {Henry}, T.~J., {Jao}, W.-C., {et~al.} 2014,
  \bibinfo{title}{{The Solar Neighborhood. XXXII. The Hydrogen Burning Limit},}
  \aj, 147, 94, \dodoi{10.1088/0004-6256/147/5/94}

\bibitem[{A. {Ginsburg} {et~al.}(2019){Ginsburg}, {Sip{\H{o}}cz}, {Brasseur},
  {Cowperthwaite}, {Craig}, {Deil}, {Guillochon}, {Guzman}, {Liedtke}, {Lian
  Lim}, {Lockhart}, {Mommert}, {Morris}, {Norman}, {Parikh}, {Persson},
  {Robitaille}, {Segovia}, {Singer}, {Tollerud}, {de Val-Borro}, {Valtchanov},
  {Woillez}, {Astroquery Collaboration}, \& {a subset of astropy
  Collaboration}}]{2019AJ....157...98G}
{Ginsburg}, A., {Sip{\H{o}}cz}, B.~M., {Brasseur}, C.~E., {et~al.} 2019,
  \bibinfo{title}{{astroquery: An Astronomical Web-querying Package in
  Python},} \aj, 157, 98, \dodoi{10.3847/1538-3881/aafc33}

\bibitem[{J.~V. {Harre} {et~al.}(2024){Harre}, {Smith}, {Barros}, {Singh},
  {Korth}, {Brandeker}, {Collier Cameron}, {Lendl}, {Wilson}, {Borsato},
  {Csizmadia}, {Cabrera}, {Parviainen}, {Correia}, {Akinsanmi}, {Rosario},
  {Leonardi}, {Serrano}, {Alibert}, {Alonso}, {Asquier}, {B{\'a}rczy}, {Barrado
  Navascues}, {Baumjohann}, {Benz}, {Billot}, {Broeg}, {Busch}, {Cubillos},
  {Davies}, {Deleuil}, {Deline}, {Delrez}, {Demangeon}, {Demory}, {Derekas},
  {Edwards}, {Ehrenreich}, {Erikson}, {Fortier}, {Fossati}, {Fridlund},
  {Gandolfi}, {Gazeas}, {Gillon}, {G{\"u}del}, {G{\"u}nther}, {Heitzmann},
  {Helling}, {Isaak}, {Kiss}, {Lam}, {Laskar}, {Lecavelier des Etangs},
  {Magrin}, {Maxted}, {Mer{\'\i}n}, {Mordasini}, {Nascimbeni}, {Olofsson},
  {Ottensamer}, {Pagano}, {Pall{\'e}}, {Peter}, {Piazza}, {Piotto}, {Pollacco},
  {Queloz}, {Ragazzoni}, {Rando}, {Rauer}, {Ribas}, {Santos}, {Scandariato},
  {S{\'e}gransan}, {Simon}, {Sousa}, {Stalport}, {Sulis}, {Szab{\'o}}, {Udry},
  {Ulmer}, {Van Grootel}, {Venturini}, {Villaver}, {Viotto}, {Walton}, {West},
  \& {Westerdorff}}]{2024A&A...692A.254H}
{Harre}, J.~V., {Smith}, A.~M.~S., {Barros}, S.~C.~C., {et~al.} 2024,
  \bibinfo{title}{{Hints of a close outer companion to the ultra-hot Jupiter
  TOI-2109 b},} \aap, 692, A254, \dodoi{10.1051/0004-6361/202451068}

\bibitem[{M. {Hippke} \& R. {Heller}(2019){Hippke} \&
  {Heller}}]{2019A&A...623A..39H}
{Hippke}, M., \& {Heller}, R. 2019, \bibinfo{title}{{Optimized transit
  detection algorithm to search for periodic transits of small planets},} \aap,
  623, A39, \dodoi{10.1051/0004-6361/201834672}

\bibitem[{D. {Huber} {et~al.}(2022){Huber}, {White}, {Metcalfe}, {Chontos},
  {Fausnaugh}, {Ho}, {Van Eylen}, {Ball}, {Basu}, {Bedding}, {Benomar},
  {Bossini}, {Breton}, {Buzasi}, {Campante}, {Chaplin},
  {Christensen-Dalsgaard}, {Cunha}, {Deal}, {Garc{\'\i}a}, {Garc{\'\i}a
  Mu{\~n}oz}, {Gehan}, {Gonz{\'a}lez-Cuesta}, {Jiang}, {Kayhan}, {Kjeldsen},
  {Lundkvist}, {Mathis}, {Mathur}, {Monteiro}, {Nsamba}, {Ong},
  {Pak{\v{s}}tien{\.{e}}}, {Serenelli}, {Silva Aguirre}, {Stassun}, {Stello},
  {Norgaard Stilling}, {Lykke Winther}, {Wu}, {Barclay}, {Daylan},
  {G{\"u}nther}, {Hermes}, {Jenkins}, {Latham}, {Levine}, {Ricker}, {Seager},
  {Shporer}, {Twicken}, {Vanderspek}, \& {Winn}}]{2022AJ....163...79H}
{Huber}, D., {White}, T.~R., {Metcalfe}, T.~S., {et~al.} 2022,
  \bibinfo{title}{{A 20 Second Cadence View of Solar-type Stars and Their
  Planets with TESS: Asteroseismology of Solar Analogs and a Recharacterization
  of {\ensuremath{\pi}} Men c},} \aj, 163, 79, \dodoi{10.3847/1538-3881/ac3000}

\bibitem[{G. {Kov{\'a}cs} {et~al.}(2002){Kov{\'a}cs}, {Zucker}, \&
  {Mazeh}}]{2002A&A...391..369K}
{Kov{\'a}cs}, G., {Zucker}, S., \& {Mazeh}, T. 2002, \bibinfo{title}{{A
  box-fitting algorithm in the search for periodic transits},} \aap, 391, 369,
  \dodoi{10.1051/0004-6361:20020802}

\bibitem[{F. {Marcadon} {et~al.}(2025){Marcadon}, {Moharana}, {Pawar}, {Pawar},
  {He{\l}miniak}, {Marques}, \& {Konacki}}]{2025arXiv250917011M}
{Marcadon}, F., {Moharana}, A., {Pawar}, T., {et~al.} 2025, \bibinfo{title}{{RX
  Gru: a short-period pre-main-sequence eclipsing binary with a distant
  circumbinary companion},} arXiv e-prints, arXiv:2509.17011,
  \dodoi{10.48550/arXiv.2509.17011}

\bibitem[{F. {Marcadon} \& A. {Pr{\v{s}}a}(2024){Marcadon} \&
  {Pr{\v{s}}a}}]{2024ApJ...976..242M}
{Marcadon}, F., \& {Pr{\v{s}}a}, A. 2024, \bibinfo{title}{{Precision Timing of
  Eclipsing Binaries from TESS Full Frame Images: Method and Performance},}
  \apj, 976, 242, \dodoi{10.3847/1538-4357/ad8571}

\bibitem[{B. {Mosser} \& T. {Appourchaux}(2009){Mosser} \&
  {Appourchaux}}]{2009A&A...508..877M}
{Mosser}, B., \& {Appourchaux}, T. 2009, \bibinfo{title}{{On detecting the
  large separation in the autocorrelation of stellar oscillation times
  series},} \aap, 508, 877, \dodoi{10.1051/0004-6361/200912944}

\bibitem[{P. {Padoan} \& {\r{A}}. {Nordlund}(2004){Padoan} \&
  {Nordlund}}]{2004ApJ...617..559P}
{Padoan}, P., \& {Nordlund}, {\r{A}}. 2004, \bibinfo{title}{{The ``Mysterious''
  Origin of Brown Dwarfs},} \apj, 617, 559, \dodoi{10.1086/345413}

\bibitem[{J.~B. {Pollack} {et~al.}(1996){Pollack}, {Hubickyj}, {Bodenheimer},
  {Lissauer}, {Podolak}, \& {Greenzweig}}]{1996Icar..124...62P}
{Pollack}, J.~B., {Hubickyj}, O., {Bodenheimer}, P., {et~al.} 1996,
  \bibinfo{title}{{Formation of the Giant Planets by Concurrent Accretion of
  Solids and Gas},} \icarus, 124, 62, \dodoi{10.1006/icar.1996.0190}

\bibitem[{R. {Rebolo} {et~al.}(1995){Rebolo}, {Zapatero Osorio}, \&
  {Mart{\'\i}n}}]{1995Nature.377..129R}
{Rebolo}, R., {Zapatero Osorio}, M.~R., \& {Mart{\'\i}n}, E.~L. 1995,
  \bibinfo{title}{{Discovery of a brown dwarf in the Pleiades star cluster},}
  \nat, 377, 129, \dodoi{10.1038/377129a0}

\bibitem[{J.~C. {Smith} {et~al.}(2012){Smith}, {Stumpe}, {Van Cleve},
  {Jenkins}, {Barclay}, {Fanelli}, {Girouard}, {Kolodziejczak}, {McCauliff},
  {Morris}, \& {Twicken}}]{2012PASP..124.1000S}
{Smith}, J.~C., {Stumpe}, M.~C., {Van Cleve}, J.~E., {et~al.} 2012,
  \bibinfo{title}{{Kepler Presearch Data Conditioning II - A Bayesian Approach
  to Systematic Error Correction},} \pasp, 124, 1000, \dodoi{10.1086/667697}

\bibitem[{K.~G. {Stassun} {et~al.}(2019){Stassun}, {Oelkers}, {Paegert},
  {Torres}, {Pepper}, {De Lee}, {Collins}, {Latham}, {Muirhead}, {Chittidi},
  {Rojas-Ayala}, {Fleming}, {Rose}, {Tenenbaum}, {Ting}, {Kane}, {Barclay},
  {Bean}, {Brassuer}, {Charbonneau}, {Ge}, {Lissauer}, {Mann}, {McLean},
  {Mullally}, {Narita}, {Plavchan}, {Ricker}, {Sasselov}, {Seager}, {Sharma},
  {Shiao}, {Sozzetti}, {Stello}, {Vanderspek}, {Wallace}, \&
  {Winn}}]{2019AJ....158..138S}
{Stassun}, K.~G., {Oelkers}, R.~J., {Paegert}, M., {et~al.} 2019,
  \bibinfo{title}{{The Revised TESS Input Catalog and Candidate Target List},}
  \aj, 158, 138, \dodoi{10.3847/1538-3881/ab3467}

\bibitem[{M.~C. {Stumpe} {et~al.}(2014){Stumpe}, {Smith}, {Catanzarite}, {Van
  Cleve}, {Jenkins}, {Twicken}, \& {Girouard}}]{2014PASP..126..100S}
{Stumpe}, M.~C., {Smith}, J.~C., {Catanzarite}, J.~H., {et~al.} 2014,
  \bibinfo{title}{{Multiscale Systematic Error Correction via Wavelet-Based
  Bandsplitting in Kepler Data},} \pasp, 126, 100, \dodoi{10.1086/674989}

\bibitem[{M.~C. {Stumpe} {et~al.}(2012){Stumpe}, {Smith}, {Van Cleve},
  {Twicken}, {Barclay}, {Fanelli}, {Girouard}, {Jenkins}, {Kolodziejczak},
  {McCauliff}, \& {Morris}}]{2012PASP..124..985S}
{Stumpe}, M.~C., {Smith}, J.~C., {Van Cleve}, J.~E., {et~al.} 2012,
  \bibinfo{title}{{Kepler Presearch Data Conditioning
  I{\textemdash}Architecture and Algorithms for Error Correction in Kepler
  Light Curves},} \pasp, 124, 985, \dodoi{10.1086/667698}

\bibitem[{N. {Unger} {et~al.}(2023){Unger}, {S{\'e}gransan}, {Barbato},
  {Delisle}, {Sahlmann}, {Holl}, \& {Udry}}]{2023A&A...680A..16U}
{Unger}, N., {S{\'e}gransan}, D., {Barbato}, D., {et~al.} 2023,
  \bibinfo{title}{{Exploring the brown dwarf desert with precision radial
  velocities and Gaia DR3 astrometric orbits},} \aap, 680, A16,
  \dodoi{10.1051/0004-6361/202347578}

\bibitem[{Z. {Zhang} {et~al.}(2024){Zhang}, {Wang}, {Ma}, {Chen}, {Wang}, {Yu},
  {Liu}, {Gao}, {Tang}, \& {Ma}}]{2024ApJS..275...32Z}
{Zhang}, Z., {Wang}, W., {Ma}, X., {et~al.} 2024, \bibinfo{title}{{Constraining
  the Presence of Companion Planets in Hot Jupiter Planetary Systems Using
  Transit-timing Variation Observations from TESS},} \apjs, 275, 32,
  \dodoi{10.3847/1538-4365/ad89a6}

\end{thebibliography}
\bibliographystyle{aasjournalv7}



\begin{longtable}{@{}lrrrr@{}}
\caption{Ephemerides for the detected variable candidates.}
\label{tab:tic}\\
\hline
TIC & \multicolumn{1}{c}{$P$}    & \multicolumn{1}{c}{$T_0$}               & \multicolumn{1}{c}{$T_{\rm mag}$} & \multicolumn{1}{c}{SDE} \\
    & \multicolumn{1}{c}{(days)} & \multicolumn{1}{c}{${\rm BJD}-2457000$} &                                                             \\
\hline
\endfirsthead
\caption{(Continued.)}\\
\hline
TIC & \multicolumn{1}{c}{$P$}    & \multicolumn{1}{c}{$T_0$}               & \multicolumn{1}{c}{$T_{\rm mag}$} & \multicolumn{1}{c}{SDE} \\
    & \multicolumn{1}{c}{(days)} & \multicolumn{1}{c}{${\rm BJD}-2457000$} &                                                             \\
\hline
\endhead
\hline
\endfoot            
5882269 & 12.8260188884 & 2474.1892892967 &     12.039 & 13.7181524935 \\
6893917 & 3.0802476261 & 2502.3874602673 &      9.021 & 181.1547448236 \\
6953740 & 3.4899918731 & 3235.7634437106 &     15.219 & 19.2327714810 \\
7020254 & 20.2730675733 & 1495.1714680932 &     12.746 & 77.9357401507 \\
7059054 & 5.1858730709 & 2503.0718053025 &     12.244 & 31.1608847155 \\
8400842 & 3.1914908447 & 1819.0732951109 &      9.441 & 153.3514705321 \\
9443323 & 4.0138990870 & 2913.8003344014 &     12.196 & 41.2088547057 \\
9725627 & 4.1567590789 & 1356.9680181768 &     10.977 & 88.0517447351 \\
14344979 & 4.1117051626 & 2259.9785001121 &     12.335 & 105.9864364451 \\
14602163 & 8.9910551979 & 3240.4397136170 &      6.709 & 40.4050025264 \\
14770122 & 3.5438319430 & 2500.3861890870 &     11.889 & 90.6610370712 \\
15932677 & 1.4367961134 & 2500.8686948662 &     10.032 & 43.3834432780 \\
15935339 & 1.7023257851 & 2474.6120733914 &     12.054 & 131.0749937723 \\
16740101 & 1.4811110319 & 1683.4474583517 &      7.582 & 246.7045640197 \\
17746821 & 3.1219939827 & 1493.1720544002 &     11.316 & 126.4315122336 \\
19028197 & 3.3366575707 & 2500.5345686303 &     10.248 & 128.1867847598 \\
20497187 & 19.5331903232 & 2516.7569361583 &     14.629 & 56.0694927580 \\
21223786 & 1.5233495274 & 2500.8952168841 &     10.855 & 21.3257206539 \\
21223810 & 3.3820609741 & 2503.5486496320 &     13.210 & 164.3769500180 \\
21337019 & 8.6579717600 & 2508.5752881042 &     11.642 & 25.3925074126 \\
22987989 & 0.7018562887 & 2500.2826660011 &     11.601 & 69.8606772477 \\
26017005 & 4.0985403046 & 2449.1144512442 &     12.549 & 94.6304507614 \\
26748546 & 5.9024973568 & 2396.3466607604 &     14.681 & 81.6746679659 \\
26826078 & 2.8753592092 & 2500.2865419284 &     10.283 & 153.5259397615 \\
27083462 & 40.9563451022 & 3343.2974366227 &     12.371 & 16.1210191207 \\
27397122 & 12.7319149310 & 1689.4611701030 &     11.748 & 76.0872849392 \\
27454154 & 1.2625439710 & 3312.8810365691 &     12.980 & 19.6496629285 \\
27685647 & 34.5846963938 & 3325.9648986518 &     10.546 & 42.4124122042 \\
27774415 & 4.9428800645 & 2393.1096397270 &     13.239 & 101.8510372889 \\
27847307 & 2.4957750049 & 2420.8086125259 &     13.676 & 83.6027555833 \\
27848472 & 1.8468337734 & 2012.1101162734 &     10.106 & 213.8703185712 \\
27916356 & 3.2347457048 & 2421.5995287674 &     12.688 & 92.3337243641 \\
28160089 & 1.7583347349 & 3368.1895972617 &     14.812 & 23.7606103411 \\
28230919 & 4.8878452384 & 1687.1968898308 &      8.508 & 166.3735823787 \\
35022727 & 3.6528387519 & 2450.6452184681 &     12.219 & 116.2000617039 \\
36592530 & 2.4841697650 & 2449.4327927980 &     10.931 & 72.8565079688 \\
37718056 & 3.6574475095 & 3212.2550806659 &     11.708 & 12.5192116995 \\
38087018 & 1.7292425684 & 2527.9236997586 &     14.603 & 70.0278077356 \\
39903405 & 5.6333988217 & 1956.2416115199 &      8.275 & 139.2828269002 \\
45179675 & 6.6228513102 & 3239.4049393183 &     14.703 & 15.9024470388 \\
45545629 & 3.2621737382 & 3237.9721348466 &     11.199 & 29.5153292634 \\
48355200 & 1.0977067754 & 3340.8423691262 &     14.240 & 11.3084740209 \\
48451130 & 30.3605511640 & 2413.7183817117 &     13.759 & 45.2499162121 \\
49428482 & 2.8268205070 & 2529.0466124006 &     11.369 & 258.8043315921 \\
51234631 & 4.1250646389 & 1846.1287741145 &     10.165 & 115.2213714024 \\
53189332 & 9.2897205767 & 1544.6478483689 &     10.990 & 71.8754591334 \\
56399553 & 3.2589005192 & 2475.9598211581 &     11.548 & 116.4763895957 \\
58731207 & 6.9030686958 & 3241.5249569028 &     14.432 & 18.2235703976 \\
58946313 & 20.5650920808 & 2512.5795081852 &     11.356 & 12.0383631160 \\
61113635 & 1.6742504259 & 3209.0967843737 &     12.010 & 14.9249099608 \\
63126950 & 31.9735663253 & 1707.7173336347 &      8.276 & 116.4604740491 \\
63189173 & 2.2984531965 & 2500.9938282505 &     12.803 & 125.0624993274 \\
68577662 & 11.1683246352 & 2478.2834380750 &      9.314 & 89.7224911989 \\
69679391 & 3.4740992578 & 1684.3157197421 &      7.552 & 329.9700052870 \\
71431780 & 13.7268375800 & 1852.7071593669 &      9.099 & 36.8943338144 \\
76419763 & 2.7884320259 & 2012.8637627560 &     11.400 & 137.6507677082 \\
81729236 & 1.0434434839 & 2501.3829564593 &     10.689 & 59.3922910773 \\
81808065 & 7.5565733049 & 3237.9719741418 &     12.261 & 20.2879871456 \\
82330323 & 0.9078147816 & 3236.0306913629 &     14.235 & 23.6849660627 \\
82928381 & 17.1522473160 & 3236.7856062830 &     14.155 & 15.6190414653 \\
83377800 & 5.7442730494 & 2476.0389898488 &     12.332 & 27.9879817273 \\
83709114 & 3.0461847042 & 2476.2792270806 &     10.412 & 41.7276728992 \\
84339983 & 2.7908255583 & 2475.6111056978 &     10.514 & 108.0334759379 \\
84453551 & 6.7496267174 & 2480.6843544651 &     10.241 & 17.2107075427 \\
84599716 & 5.3191460868 & 2476.1063523121 &     11.789 & 57.1830612421 \\
86396382 & 1.0914115799 & 1843.0079914354 &     11.097 & 189.6979331819 \\
91987762 & 45.5212960100 & 1894.2636394497 &      7.410 & 54.1317113785 \\
97735908 & 4.1137593641 & 2475.4274282968 &      8.224 & 156.6689539258 \\
99834717 & 1.7063390505 & 1871.7177616896 &     12.212 & 140.3364445399 \\
103751498 & 3.7146838799 & 2392.0944447307 &     10.910 & 126.4522558370 \\
113921235 & 143.6320375761 & 2518.4010572771 &     11.328 & 20.7384146478 \\
115524421 & 8.8047118596 & 2700.6346026504 &     11.591 & 23.2424210977 \\
120043638 & 2.7036121674 & 2391.3850891406 &     13.622 & 63.6396488202 \\
120254134 & 9.3305336839 & 3312.8979271831 &     14.585 & 9.7387322067 \\
120255950 & 31.5795420992 & 2399.6293453276 &     10.611 & 43.1226863355 \\
120757718 & 3.0300587688 & 1683.3569316636 &     11.025 & 207.0793285856 \\
121019265 & 12.7715090381 & 3319.3958479966 &     12.347 & 13.1580211684 \\
121022334 & 48.5861699373 & 2429.5128224829 &     14.381 & 14.5038970878 \\
121277143 & 21.5133155647 & 1687.2204368156 &      9.997 & 60.1575603761 \\
121393780 & 22.9138720699 & 1685.6072709603 &     11.461 & 258.7737730590 \\
121598562 & 3.3502720714 & 1683.6083794781 &     11.066 & 205.7251028286 \\
121660779 & 27.1956505125 & 3313.9171639587 &     13.464 & 12.7349568487 \\
121660904 & 4.8854369936 & 2393.5329934866 &     12.453 & 82.9271746967 \\
121865891 & 1.2371118083 & 3313.8534928494 &     13.350 & 31.4043101648 \\
121941586 & 12.0097247048 & 3318.7963501168 &     12.870 & 25.9167437525 \\
122069957 & 26.8216807828 & 3326.8014334085 &     12.042 & 15.7014412778 \\
122226990 & 24.9559851641 & 1700.5405399033 &     11.290 & 35.1537862235 \\
122304494 & 2.9760346355 & 1685.9798078344 &     10.855 & 347.4168818818 \\
122304858 & 17.7341914255 & 2392.6889003740 &     14.332 & 25.9169548386 \\
122304887 & 8.8645489546 & 3314.8791553821 &     12.068 & 13.6254463387 \\
122446960 & 18.6120499683 & 1689.2554939929 &     13.057 & 146.0416570712 \\
123233041 & 3.5225167418 & 2392.4639657547 &     13.145 & 78.5077644591 \\
123315911 & 2.8085912905 & 3314.6914582712 &     13.954 & 13.1586953367 \\
123316518 & 1.6084195752 & 3314.0202840510 &     13.249 & 22.3892753981 \\
123317836 & 8.7786273904 & 3318.3150449206 &     15.089 & 15.6586726980 \\
123417316 & 31.7645226822 & 2419.7423487186 &     12.950 & 13.9021750975 \\
123445835 & 10.0367776870 & 3321.8401475826 &     14.079 & 15.2880046453 \\
123445947 & 3.1045778297 & 3313.6009370480 &     14.520 & 20.3691177102 \\
123495874 & 7.8915939137 & 2397.8861790726 &     12.993 & 79.9402478458 \\
137347686 & 0.6810785733 & 3312.8738229381 &     12.084 & 46.2291907759 \\
137549549 & 43.9342174067 & 2403.2913651789 &     14.792 & 11.6660336302 \\
137881699 & 2.9721218229 & 1872.6358794889 &     12.512 & 48.2152928051 \\
137893328 & 1.6065993327 & 3313.4814343037 &     13.765 & 32.2301078223 \\
137902578 & 3.7785858666 & 3313.1616551942 &     14.054 & 16.6268007209 \\
138093387 & 26.7476066231 & 3327.2697715786 &     12.211 & 30.9346646904 \\
138100436 & 4.5547992174 & 3314.0826935012 &     11.050 & 31.1523264173 \\
138168780 & 3.7649692201 & 1817.5887392895 &      9.883 & 229.8083639119 \\
138294130 & 3.0564649333 & 1955.9261761110 &     10.651 & 78.2862567641 \\
138643306 & 4.5339527433 & 3315.4096388756 &     13.673 & 17.6353678153 \\
138888759 & 115.8542688500 & 2414.6146660815 &     11.172 & 20.7807793362 \\
138968089 & 42.9452153662 & 1723.6559372738 &     11.646 & 13.8973597665 \\
138970037 & 19.0582662946 & 3322.0940556953 &     13.564 & 18.2617742924 \\
142381532 & 38.8135383273 & 1715.1892670714 &      9.772 & 76.6798223364 \\
142387023 & 22.8089240232 & 1879.5370264755 &      8.975 & 20.7012163129 \\
142394656 & 8.1576687053 & 1711.7535929538 &     10.149 & 92.6368714485 \\
144700903 & 2.3266645710 & 1470.5700212865 &     12.678 & 72.2438300718 \\
147797743 & 3.6916057393 & 2582.2876523847 &     10.745 & 144.5647761969 \\
147950620 & 2.3106609786 & 1684.9186064520 &     10.478 & 178.0985473363 \\
149918151 & 1.5407890247 & 2420.3128545843 &     12.918 & 138.2371441037 \\
157586003 & 3.3325818225 & 2011.2728883189 &     12.055 & 95.3218775618 \\
158002130 & 9.6867766083 & 1691.0459192600 &     10.112 & 97.6900861995 \\
158172624 & 1.8279906102 & 3313.5524771676 &     13.815 & 23.1922672610 \\
158316612 & 18.7948500438 & 1693.1517056356 &     13.431 & 61.7608834406 \\
158324245 & 1.7635933286 & 1683.5514393794 &     10.231 & 164.3317424385 \\
158388163 & 4.2253287862 & 2394.5438145916 &     13.458 & 92.3192660183 \\
158490401 & 12.7139492176 & 2394.3261855892 &     11.034 & 178.3941524320 \\
158844163 & 1.0620221119 & 3313.6867372723 &      9.854 & 34.7896897498 \\
158984540 & 15.2453719118 & 3313.9420585412 &     13.158 & 21.9879220899 \\
159098316 & 3.5788427733 & 2393.9722644298 &     13.423 & 92.2942354826 \\
159440299 & 7.9354023312 & 3340.5508702204 &     14.176 & 11.2157946825 \\
159716878 & 70.5842283093 & 1693.7965410505 &     14.146 & 61.5903501915 \\
163539739 & 14.4750923739 & 1711.9613518603 &     11.782 & 33.2874014678 \\
164457525 & 27.3219701731 & 1708.5423874395 &     13.033 & 84.0621715669 \\
164558072 & 1.2137385704 & 1683.6725386480 &     10.510 & 208.5896432342 \\
164658882 & 26.6372986692 & 3315.7826557076 &     13.511 & 12.8185049833 \\
164883843 & 1.1876118035 & 3313.1061718748 &     12.813 & 23.7523439580 \\
164998842 & 26.0584282153 & 3316.4714967242 &     13.511 & 12.3758431132 \\
168699373 & 4.3012181572 & 1931.4287337106 &     12.447 & 89.7561996024 \\
169226822 & 4.1781021798 & 1548.1117347544 &      9.630 & 161.4583969231 \\
169458450 & 6.9322104602 & 1683.5941967565 &     11.367 & 58.8691908630 \\
169459587 & 17.0911152471 & 3322.5004175506 &     12.865 & 13.0663191313 \\
169459899 & 3.0412386443 & 3314.0198438418 &     13.296 & 14.0008898517 \\
169461816 & 17.8552055724 & 1693.4984446200 &     10.931 & 48.9443051858 \\
169819162 & 17.7452692355 & 1697.2569456473 &     12.167 & 160.3848111986 \\
169819920 & 41.6438318244 & 2435.8521211772 &     11.609 & 21.7064098462 \\
170344769 & 11.2587927802 & 1694.4116364403 &     13.111 & 92.6085273598 \\
171027701 & 0.6561373176 & 3340.0464186295 &     14.704 & 16.4563108232 \\
171505317 & 2.4458275855 & 3340.0565100699 &     17.267 & 12.8007819948 \\
171972797 & 2.0069226446 & 3314.5976766269 &     14.147 & 15.6327039527 \\
171975104 & 1.4130329556 & 3340.7672912388 &     14.550 & 14.2262066728 \\
172370679 & 29.0904622805 & 1711.9542764403 &     12.583 & 61.1414341949 \\
172422394 & 2.4043762591 & 1683.3987287248 &     11.547 & 379.9393539988 \\
172422658 & 9.0652779196 & 3319.5112322685 &     11.624 & 14.4588809289 \\
173103335 & 55.8314640494 & 2558.3519459011 &     10.546 & 9.7838129037 \\
176220787 & 5.2152637505 & 2092.5330891775 &      9.483 & 76.4278245537 \\
178217113 & 2.9956619511 & 2501.9091636339 &     11.051 & 109.0317179232 \\
184091846 & 3.4802485046 & 3314.1893649512 &     13.564 & 18.2913163468 \\
184167572 & 16.3801622022 & 3323.2717787585 &     13.176 & 16.5011591872 \\
184892124 & 1.9827507252 & 2500.6737884522 &     16.057 & 70.4779290499 \\
185059488 & 0.7397664860 & 1683.7960484052 &     11.219 & 236.5796768543 \\
185060842 & 0.8248478398 & 1683.8896164889 &     11.868 & 218.0568017688 \\
186812530 & 3.6773580116 & 1492.0530246741 &      9.198 & 167.5563988335 \\
186817940 & 6.3529370881 & 3235.9577001441 &      9.489 & 16.4846630137 \\
188589164 & 2.6156254437 & 1956.4032447046 &     10.271 & 27.3454996077 \\
195192253 & 3.4758874931 & 3264.4412195571 &     13.634 & 15.1535500589 \\
198108326 & 3.2130538393 & 1933.5433507806 &     11.755 & 140.6254405205 \\
198153540 & 89.1795148706 & 1762.9685296294 &     10.519 & 16.4461722501 \\
198588220 & 4.8101711343 & 1958.0728819899 &      9.633 & 122.6516399122 \\
199376584 & 18.7120172605 & 1692.2516404517 &      8.042 & 81.5859440436 \\
199671901 & 4.4570717622 & 2829.1353745548 &     10.973 & 62.3593433139 \\
203189770 & 3.6127185513 & 2420.2491779903 &     10.422 & 120.7636406800 \\
203287192 & 4.4454976111 & 2504.1456141267 &     12.148 & 58.9653595186 \\
207468071 & 20.3806045125 & 1933.1735605231 &      9.294 & 66.5231901046 \\
219015370 & 3.9153114239 & 2393.0032829624 &     10.946 & 99.8606228416 \\
219854185 & 3.9444213798 & 1930.7440627097 &     10.804 & 121.6682762012 \\
219854519 & 31.0339468112 & 2394.9416130727 &     10.548 & 62.2124735805 \\
229400092 & 4.1420154580 & 1933.3248476611 &     11.209 & 135.7836213259 \\
229510866 & 2.1032108954 & 1957.8157867963 &     10.079 & 144.0490023530 \\
229791084 & 1.6453144193 & 1957.0486699147 &     12.286 & 148.8257851676 \\
232612416 & 4.3601799887 & 1687.1170422090 &     11.115 & 153.3526936240 \\
232967440 & 7.0643372120 & 1688.7199054348 &     10.294 & 147.3287005352 \\
233948455 & 1.4822442880 & 2420.1214130554 &     12.883 & 130.7242506992 \\
236387002 & 7.2009793133 & 1958.6748602303 &     10.398 & 153.7097099474 \\
236714379 & 1.0180234662 & 1683.7408785059 &     10.794 & 169.2975132611 \\
236815160 & 2.5345822438 & 2391.5514683638 &     10.797 & 84.2217332327 \\
236887394 & 1.4200258529 & 1766.0056116562 &     11.848 & 229.8230446756 \\
237104103 & 4.5371033016 & 1934.1275620535 &     10.959 & 102.0828424525 \\
239289191 & 2.6180029202 & 3315.1090105980 &     12.885 & 17.4360532529 \\
239816546 & 2.9887690815 & 2500.7404243547 &     11.977 & 110.0354182671 \\
240193270 & 3.6719221769 & 3314.7006665726 &     13.278 & 18.8826937232 \\
242940281 & 7.1945631989 & 3213.9154321515 &     12.804 & 22.9598355784 \\
243244680 & 19.4915862006 & 2517.3789605987 &      9.669 & 43.8046630270 \\
248382917 & 0.8760481092 & 3208.6523988031 &     11.899 & 10.3199626653 \\
248387177 & 44.3570281297 & 2132.8168676096 &     11.242 & 31.1577583176 \\
248942621 & 17.0932809784 & 2125.7423960338 &     12.078 & 46.3642255039 \\
252479260 & 3.2122189794 & 1871.6334586741 &      9.007 & 243.0694713952 \\
255930614 & 2.8758667835 & 1957.5568018801 &      7.626 & 135.9329956236 \\
256722647 & 2.9222273498 & 2720.7280918721 &     11.883 & 36.5054961882 \\
257060897 & 3.6600984181 & 2393.4176176516 &     11.263 & 93.3308543085 \\
257217518 & 37.2490691026 & 2533.8925865147 &     11.555 & 37.7684665970 \\
257395093 & 14.4249208656 & 2451.2750866657 &     12.286 & 23.6429589108 \\
257558789 & 0.6717850415 & 3209.3991539192 &     11.503 & 18.4471895919 \\
257816591 & 50.7353117215 & 2498.2694408187 &     13.550 & 10.5607127384 \\
258920431 & 5.9838356497 & 2390.7618798569 &     11.749 & 74.9305179869 \\
259506033 & 1.9023348661 & 2913.3164305106 &     12.382 & 69.6697506833 \\
262662119 & 4.5335244480 & 2450.1209265810 &     12.180 & 39.4538106891 \\
266012991 & 12.6834262622 & 2460.2700040367 &     11.243 & 14.8554708342 \\
266015496 & 0.9289447628 & 3209.2389157629 &      8.382 & 10.9636602746 \\
268074192 & 18.7297027741 & 3208.9180635102 &      8.511 & 18.5680852071 \\
268166705 & 9.9424912711 & 2420.9525155180 &     13.533 & 78.2803497885 \\
268289462 & 1.7525699717 & 1684.6583194484 &     10.880 & 386.2790052541 \\
268299560 & 8.7530290467 & 1691.6745941310 &     11.614 & 165.0474490346 \\
268305489 & 1.2622702156 & 1683.7030275792 &     10.809 & 148.5633536714 \\
268380075 & 0.8219878054 & 3313.6152599171 &     12.934 & 26.1170087940 \\
268381905 & 21.0248792986 & 3321.6868225633 &     12.628 & 43.1097875826 \\
268403451 & 2.8641564479 & 1985.7927114554 &     10.052 & 174.5827121873 \\
268603980 & 20.3594862740 & 3312.9142005904 &     13.766 & 9.9242406882 \\
268704157 & 6.8107962034 & 2423.3292374167 &     12.251 & 75.6657332075 \\
268711054 & 14.3567508844 & 3313.4164751692 &     13.335 & 13.7426998151 \\
268823307 & 3.5484769609 & 2423.4190188529 &     12.765 & 79.7366341925 \\
269029623 & 1.3435836487 & 3313.1258964095 &     12.438 & 28.2590977129 \\
269701147 & 8.8803231518 & 1715.3550673123 &      8.293 & 123.0269926169 \\
270380593 & 3.8361966059 & 1414.1263025385 &     10.731 & 131.1854877261 \\
270468559 & 4.6418477487 & 1519.4622650819 &     11.560 & 108.0914099957 \\
270604417 & 7.5428406594 & 2233.3320366859 &     10.816 & 46.3911357860 \\
270616245 & 2.1798021240 & 3314.6682483666 &     13.467 & 48.2501183626 \\
270700608 & 23.8760316024 & 1688.8039983232 &     11.315 & 48.3486688856 \\
270864283 & 25.8807991569 & 3327.7672137414 &     13.100 & 25.9593440558 \\
270957624 & 2.7434553137 & 3314.8210142114 &     13.224 & 14.1485853329 \\
271040768 & 10.3671051625 & 1687.5897413199 &     15.214 & 71.0911480125 \\
271355638 & 8.1849152622 & 3316.5049030210 &     13.877 & 20.0941330000 \\
271548206 & 7.4484402813 & 1687.4270072754 &     14.671 & 119.9683687595 \\
271663183 & 1.0248357817 & 3340.7419577972 &     14.180 & 19.1385688378 \\
271759162 & 25.6001612360 & 3319.3906553417 &     13.784 & 19.9494059735 \\
271761549 & 0.6916837605 & 3313.4656112242 &     12.843 & 22.1378825641 \\
272071992 & 24.9834754469 & 3335.2398230387 &     13.001 & 19.8610545007 \\
272173724 & 27.3088945219 & 3333.1203422182 &     13.718 & 17.1645361438 \\
272363671 & 11.3729971205 & 3321.4451381706 &     13.998 & 23.3224243748 \\
272369124 & 27.7953266916 & 1693.8130491601 &     14.339 & 74.5142587617 \\
272375892 & 2.5381087866 & 3314.5279101851 &     12.430 & 19.8189006801 \\
272487801 & 14.1495178471 & 3318.4046270855 &     12.371 & 16.8666880004 \\
272487856 & 23.8399864216 & 3322.4364594969 &     13.094 & 21.1173894590 \\
272489100 & 1.7905209301 & 1684.8432598835 &     12.052 & 270.2722525907 \\
272490653 & 1.1783381942 & 3313.0588772968 &     13.137 & 25.7719895887 \\
272707711 & 2.6889321492 & 2420.9556699468 &     15.089 & 28.5357265673 \\
272719652 & 80.3837750248 & 1689.4127966317 &     11.236 & 43.1375020788 \\
273042650 & 6.1502359259 & 1688.3270803207 &     10.476 & 98.2921696234 \\
273131564 & 7.8633513977 & 1684.0380914052 &     12.016 & 113.9541955044 \\
273376048 & 25.2853603432 & 1685.8906184717 &     11.718 & 173.7307492737 \\
273505466 & 2.1551933441 & 3313.8218211029 &     13.470 & 22.7112335500 \\
273874849 & 1.4857301400 & 2421.4334214378 &     13.641 & 52.6237677761 \\
274030932 & 1.0565171234 & 3313.7522157534 &     12.891 & 20.8698822557 \\
275489982 & 13.7876289224 & 1692.3877087202 &     11.717 & 122.2598961175 \\
275570163 & 1.1523200655 & 3313.7194044085 &     13.237 & 14.9524112335 \\
275570329 & 2.1446402070 & 2391.3617892980 &     13.073 & 79.9218773393 \\
275575525 & 2.4040389906 & 3314.8271554545 &     13.060 & 25.1243350604 \\
281731203 & 11.6334716449 & 1546.6815936282 &      9.962 & 72.4340360897 \\
281885301 & 7.9204601976 & 2530.2783575618 &     12.305 & 37.1090627398 \\
284326455 & 3.7998635600 & 2475.9805516044 &     12.234 & 134.5493317002 \\
284475976 & 2.1436154363 & 1931.5715175196 &     11.154 & 147.0713855921 \\
285048486 & 3.4914455566 & 1843.2666498167 &     10.792 & 76.9000101897 \\
286923464 & 6.1349187149 & 1712.6667933306 &      7.456 & 130.2168464495 \\
288735205 & 3.4780044699 & 1686.6946785012 &     11.153 & 246.8617045504 \\
293612446 & 2.6267314897 & 2176.1088240397 &     12.036 & 74.7919406212 \\
293687315 & 2.8717335383 & 1932.4151169334 &      9.823 & 148.7281531939 \\
296323861 & 0.8467133168 & 3340.4653663177 &     12.137 & 38.7160776335 \\
298964574 & 26.6371582051 & 3315.8189341653 &     11.841 & 14.2049921714 \\
299096355 & 41.0766085023 & 1703.7402765456 &     10.944 & 271.3558528524 \\
301289516 & 1.2089595744 & 2448.3193904377 &      9.068 & 24.2064539541 \\
302773669 & 21.2160062688 & 1809.0846909964 &      7.635 & 138.7154276048 \\
303568322 & 0.9835465241 & 3340.7579751680 &     14.547 & 27.6443609151 \\
306263608 & 20.7727942584 & 1767.4238713570 &      8.572 & 34.6970952606 \\
307016006 & 3.2945942487 & 1471.3504160013 &     13.789 & 48.9100632431 \\
307809773 & 2.7694655131 & 2475.7575658393 &      7.899 & 56.8429860770 \\
311035838 & 2.8997465918 & 1739.4492951881 &     10.716 & 164.5801808431 \\
311133118 & 3.5071578012 & 2666.8566865055 &     10.532 & 26.8173755373 \\
316470945 & 12.8960375243 & 1777.2476470350 &      8.873 & 11.9050882238 \\
321668398 & 5.6441292468 & 1961.3344699357 &     11.543 & 103.2906604077 \\
325315305 & 10.3884666594 & 3264.8837802178 &     12.555 & 20.7311668210 \\
327952677 & 3.3969565353 & 2090.1410244339 &     12.141 & 75.1820720069 \\
330687113 & 5.1138070955 & 2503.1237619566 &     10.999 & 51.4233200506 \\
332064670 & 0.7365411275 & 1870.6974002288 &      5.206 & 122.5440851923 \\
336128819 & 7.8519336698 & 2017.7054726695 &     11.737 & 40.1352155219 \\
336893387 & 26.8269653309 & 2471.6718645903 &     11.878 & 9.4264898646 \\
336956423 & 19.0242828088 & 2466.6040051762 &     10.634 & 9.1811559886 \\
337079474 & 13.3465675372 & 2475.6152697665 &     11.171 & 69.0699488545 \\
343019899 & 28.5586396750 & 2837.5301288868 &     10.062 & 21.4089525982 \\
347013211 & 9.0823241892 & 2854.8852546949 &     10.780 & 16.3181887341 \\
348667759 & 14.0924124569 & 3211.1632056146 &     12.181 & 21.7588197707 \\
349827430 & 5.5515458863 & 1684.8096711563 &      7.918 & 230.6908181370 \\
350132371 & 1.0318217431 & 1739.1327934369 &      9.258 & 50.6179022111 \\
350987976 & 0.6487755381 & 1683.5304394414 &      9.775 & 171.2086574645 \\
351055067 & 28.7992743154 & 2419.3338374024 &     14.053 & 12.0244031979 \\
351193098 & 35.8999452580 & 3339.6612721100 &     13.393 & 39.9322390753 \\
351803550 & 7.2296353291 & 3317.1757901861 &     12.325 & 12.9928968192 \\
351804693 & 24.6167606269 & 3321.6950511378 &     13.552 & 29.2070579526 \\
355867695 & 3.1274909335 & 1684.0047113410 &     10.812 & 35.5958298439 \\
360028729 & 2.2436252287 & 3313.7927315362 &     12.297 & 52.1322088221 \\
363445389 & 5.2811966513 & 2526.9053242292 &     11.057 & 94.7180051686 \\
363548415 & 26.4220367200 & 2526.5944346304 &     11.919 & 12.5045792406 \\
365003901 & 1.7276547798 & 3209.9666797614 &     15.087 & 19.4433827258 \\
365733349 & 2.6998175627 & 1712.3622273360 &      9.931 & 151.6457352972 \\
366443576 & 47.4228221699 & 2267.8536798872 &     11.568 & 13.6146765512 \\
366622912 & 3.1149059868 & 1493.2811331558 &     13.799 & 125.4436357099 \\
367191694 & 0.6683839340 & 2500.5753042691 &      9.518 & 120.3437130764 \\
367366318 & 2.7347974180 & 1816.5093861033 &      8.149 & 244.8598598130 \\
367858035 & 20.7053231620 & 1806.6907917345 &     10.091 & 65.7418811259 \\
369478173 & 6.8838831795 & 3213.8063446142 &     13.212 & 14.6599381435 \\
373693175 & 1.3273591336 & 1900.7984182078 &     11.628 & 123.7411818015 \\
375506058 & 2.6502180932 & 1712.6796187056 &      7.798 & 311.0880516994 \\
376939759 & 5.8176146525 & 2452.3176826649 &     13.012 & 46.0223489362 \\
376981340 & 4.0248289787 & 2448.1116663290 &     12.813 & 74.6273531914 \\
377064495 & 10.7790086311 & 1527.0452060634 &      9.527 & 49.1832425757 \\
378014055 & 1.3337727675 & 3313.7820592031 &     14.887 & 21.6778835729 \\
378082998 & 20.3596837727 & 3312.9125430823 &     15.159 & 9.6684790097 \\
380619414 & 2.6556223165 & 2527.0530856706 &     10.150 & 179.1750464209 \\
380907135 & 4.0461133320 & 2450.1138702010 &     10.556 & 76.6472707672 \\
386890582 & 1.7445436413 & 2500.8767502056 &     12.420 & 277.7183090801 \\
386891333 & 19.6405168380 & 2509.2041657422 &     10.725 & 49.5516994208 \\
386908047 & 17.0347967966 & 2509.6987975641 &     11.483 & 72.8023843809 \\
387141176 & 2.9323161899 & 3236.9676799241 &     14.976 & 18.8669304449 \\
387690507 & 6.3875110820 & 1470.2136327811 &     13.529 & 90.9363021612 \\
387716346 & 1.6961541597 & 3235.9208628383 &     14.158 & 19.1250549998 \\
388777783 & 15.9603517760 & 2460.6553322590 &     13.239 & 35.4190156488 \\
394050135 & 8.8722175794 & 2391.1780865395 &     10.946 & 77.7324903969 \\
394177355 & 8.6530935386 & 1690.4874811745 &      7.748 & 93.6447130001 \\
396562848 & 1.9495302593 & 1792.6052892856 &     11.055 & 67.5778029464 \\
398572544 & 3.4087820477 & 2449.1892854468 &     11.624 & 69.8680134816 \\
399722652 & 26.3067720671 & 2130.0322950065 &     11.833 & 24.2729861590 \\
399860444 & 2.4706059294 & 2012.5132302563 &     10.854 & 210.0545296265 \\
405687343 & 1.0661213423 & 1683.4119639720 &      7.619 & 68.6392643670 \\
405717754 & 4.4380635770 & 2393.1460126336 &     12.961 & 85.5594052953 \\
408636441 & 18.8499695288 & 1745.4689196954 &      9.928 & 79.5452380862 \\
411839167 & 6.1802111886 & 2175.3292804094 &     10.824 & 29.3359708978 \\
414920805 & 2.3104813471 & 3209.3956348864 &     11.045 & 35.6552521388 \\
416280178 & 2.7092251130 & 3285.9819440880 &     12.725 & 26.1865139553 \\
416284206 & 2.3804271822 & 3313.3658309747 &     13.738 & 21.5949540424 \\
417646390 & 3.8017276997 & 2720.3729805999 &     10.856 & 77.6795403002 \\
417678531 & 0.8660195974 & 3367.7831357017 &     15.720 & 20.7758422015 \\
417948359 & 3.3159686606 & 1713.0336643534 &     11.024 & 50.3793427844 \\
422655579 & 2.9036698215 & 1413.1457745229 &     10.164 & 108.2807565419 \\
424435940 & 5.0171813638 & 1986.9844471070 &     11.231 & 143.7375059926 \\
424865156 & 2.2047570418 & 1684.7655947359 &     10.027 & 168.7534826694 \\
427740976 & 2.3494504907 & 2475.9924696144 &     11.451 & 93.6070198842 \\
427761355 & 1.9021153336 & 2854.4229272088 &      8.751 & 45.4687745559 \\
428679607 & 350.3111182046 & 1891.0952580375 &      9.529 & 20.7081806873 \\
430049349 & 2.4700727021 & 2474.3948730916 &      9.770 & 158.9868718679 \\
434133200 & 14.1461035320 & 2450.8489831795 &     11.841 & 36.3698050694 \\
434165964 & 3.7342523763 & 2449.2712591335 &     11.974 & 76.5439709480 \\
435880578 & 0.9699374577 & 2474.4501914719 &     11.939 & 103.8953887573 \\
435907158 & 1.4282664931 & 2475.5648714507 &     11.089 & 20.0785205618 \\
437064171 & 0.9583640680 & 3262.9586954568 &     12.170 & 19.2156501984 \\
438071843 & 2.6207271964 & 1470.5468630760 &     11.077 & 95.9072296924 \\
438102256 & 4.4481098726 & 2476.2993456050 &     11.210 & 187.1676148693 \\
438338723 & 1.8489767375 & 2201.7641806729 &     11.935 & 199.6123530166 \\
440727989 & 1.1829404962 & 2501.0504514445 &     10.866 & 157.3953118922 \\
440777904 & 3.3552153041 & 1495.0595685428 &     11.307 & 145.4860798731 \\
441739020 & 40.7506702492 & 1750.8711104772 &      9.468 & 127.2142840634 \\
444013020 & 2.3114557789 & 2231.0997908372 &     11.302 & 150.3591645803 \\
445805961 & 24.2838072610 & 1836.9514541919 &      8.913 & 69.9564165270 \\
446175401 & 0.7282794901 & 1738.9569076038 &     10.857 & 61.6547162019 \\
453064665 & 2.6453016373 & 2911.8522863504 &     11.889 & 55.0824097964 \\
453211454 & 9.0589171604 & 1494.4345192838 &      7.926 & 31.2556607393 \\
458478250 & 2.2552932687 & 1683.9098296146 &      9.767 & 352.0169691473 \\
467179528 & 10.8948906398 & 1691.0016813155 &     11.040 & 97.3132709004 \\
468987719 & 3.3327098233 & 1493.5418879187 &     12.766 & 118.9533921801 \\
\end{longtable}

\end{document}